\documentclass[aps,prl,
 amsmath,amssymb,
 twocolumn, groupedaddress,
 showpacs]{revtex4-1}
\usepackage{bm}
\usepackage{graphicx}
\usepackage{gensymb}
\usepackage{amsmath}
\usepackage{amssymb}
\usepackage{ulem}

\begin{document}

\title{Unexpected Dirac-Node Arc in the Topological Line-Node Semimetal HfSiS}
\author{D. Takane,$^1$ Z. Wang,$^2$ S. Souma,$^{3,4}$ K. Nakayama,$^1$ C. X. Trang,$^1$ T. Sato,$^{1,4}$ T. Takahashi,$^{1,3,4}$ and Yoichi Ando$^2$}

\affiliation{$^1$Department of Physics, Tohoku University, Sendai 980-8578, Japan\\
$^2$Institute of Physics II, University of Cologne, K$\ddot{o}$ln 50937, Germany\\
$^3$WPI Research Center, Advanced Institute for Materials Research, Tohoku University, Sendai 980-8577, Japan\\
$^4$Center for Spintronics Research Network, Tohoku University, Sendai 980-8577, Japan
}

\date{\today}

\begin{abstract}
    We have performed angle-resolved photoemission spectroscopy on HfSiS, which has been predicted to be a topological line-node semimetal with square Si lattice. We found a quasi-two-dimensional Fermi surface hosting bulk nodal lines, alongside the surface states at the Brillouin-zone corner exhibiting a sizable Rashba splitting and band-mass renormalization due to many-body interactions. Most notably, we discovered an unexpected Dirac-like dispersion extending one-dimensionally in {\it k} space -- the Dirac-node arc -- near the bulk node at the zone diagonal. These novel Dirac states reside on the surface and could be related to hybridizations of bulk states, but currently we have no explanation for its origin. This discovery poses an intriguing challenge to the theoretical understanding of topological line-node semimetals.
\end{abstract}


\maketitle

Topological insulators (TIs) realize a novel state of matter where an insulating bulk with an inverted energy gap induced by strong spin-orbit coupling (SOC) is accompanied by gapless edge or surface states (SSs) protected by the time-reversal symmetry \cite{AndoReview, ZhangReview, HasanReview}. The discovery of TIs triggered the search for new types of topological materials based on other symmetries, as represented by topological crystalline insulators (TCIs) where gapless SSs are protected by space-group symmetry (specifically mirror symmetry) of the crystal \cite{FuPRL2011, HsiehNC2012, TanakaNP2012, XuNC2012, DziawaNM2012}. Topological semimetals are recently becoming a leading platform for realizing such novel topological states.

   In contrast to conventional semimetals with a finite band-overlap between valence band (VB) and conduction band (CB), topological semimetals are categorized by the band-contacting nature between VB and CB in the Brillouin zone (BZ); point-contact (Dirac/Weyl semimetals) or line-contact (line-node semimetals; LNSMs). The existence of three-dimensional (3D) Dirac semimetals was first confirmed by angle-resolved photoemission spectroscopy (ARPES) of Cd$_3$As$_2$ \cite{HasanNC2014, BorisenkoPRL2014} and Na$_3$Bi \cite{ShenScience2014}, where the VB and CB contact each other at the point (Dirac point) protected by rotational symmetry of the crystal \cite{WangPRB2012, WangPRB2013}. Recent ARPES studies on noncentrosymmetric transition-metal monopnictides \cite{XuScience2015, LvPRX2015, YangNP2015, SoumaPRB2016} have clarified pairs of bulk Dirac-cone bands and Fermi-arc SSs, supporting their Weyl-semimetallic nature \cite{WengPRX2015, HuangNC2015}. While the existence of Weyl semimetals with point nodes has been confirmed experimentally, the experimental studies of LNSMs with line nodes are relatively scarce \cite{PbTaSe2Hasan, AstNC2016, HasanPRB2016, AdamNC2016} despite many theoretical predictions \cite{ZhengArXiv2015, YuPRL2015, KimPRL2015, FangPRB2015, BianArXiv2015, XiaAPLMat2015, YamakageJPSJ2016}.

  Recently, it was theoretically proposed by Xu {\it et al.} that ZrSiO with PbFCl-type crystal structure (space group $P$4$/nmm$) and its isostructural family WHM (W = Zr, Hf, or La; H = Si, Ge, Sn, or Sb; M = O, S, Se and Te; see Fig. 1(a) for crystal structure) may host the LNSM phase protected by glide-mirror symmetry of the crystal \cite{XuPRB2015}. Subsequent ARPES studies on ZrSiS \cite{AstNC2016, HasanPRB2016} and ZrSnTe \cite{ZrSnTeHong} observed the symmetry-protected band crossing at $\bar{X}$ [see Fig. 1(b)] as well as the diamond-shaped Fermi surface (FS) hosting line nodes, in good agreement with the band calculations \cite{XuPRB2015}. These studies demonstrated the realization of LNSM phase as well as an appearance of nearly-flat  SSs around the $\bar{X}$  point, both were explained on the basis of band calculations.
  
  In this Letter, we report the ARPES results on HfSiS. In addition to the overall VB structure which is in support of the LNSM nature of HfSiS, we found new spectral features, such as a large Rashba splitting of SSs at $\bar{X}$, a dispersion kink at $\sim$0.05 eV, and most importantly, unexpected Dirac-like SSs forming a ``Dirac-node arc''. This is a rare case in the research of topological materials that experiment finds novel SSs that were not predicted by theory.

\begin{figure}
\begin{center}
 \includegraphics[width=3.3in]{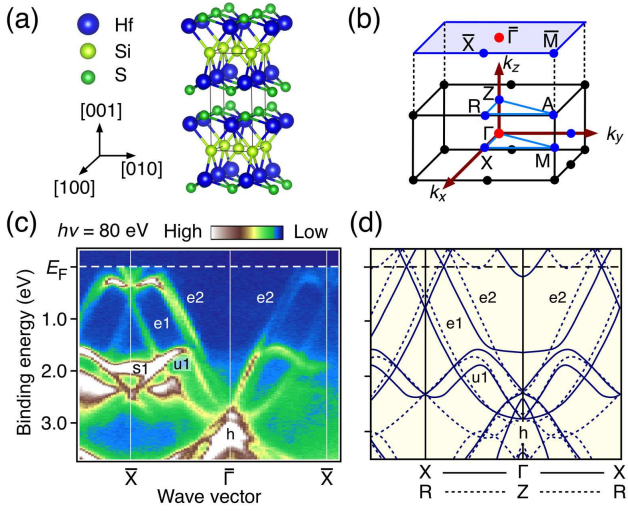}
  \hspace{-0.2in}
\caption{(color online). (a) Crystal structure of HfSiS. (b) Bulk tetragonal BZ and corresponding surface BZ (blue). (c) Plot of ARPES intensity in the VB region as a function of $k_x$ and $E_{\rm B}$ measured along the $\bar{\Gamma}\bar{X}$ cut at $h\nu$ = 80 eV. (d) Calculated bulk-band dispersion \cite{XuPRB2015} along the ${\Gamma}X$ ($k_z$=0; solid curves) and $ZR$ ($k_z$=$\pi$; dashed curves) cuts.}
\end{center}
\end{figure}

Figure 1(c) shows the ARPES-intensity plot in the VB region as a function of wave vector and binding energy ($E_{\rm B}$) measured along the $\bar{\Gamma}\bar{X}$ cut at $h\nu$ = 80 eV (see Supplemental Materials for details of sample preparation \cite{ZhiweiAPL} and ARPES measurements). One can notice several dispersive bands; holelike bands at $\bar{\Gamma}$ (h) with the top of dispersion at $E_{\rm B}${$\sim$}3 eV, electronlike bands centered at $\bar{\Gamma}$ (e1, e2), and an undulating band (u1) with the top of dispersion midway between $\bar{\Gamma}$ and $\bar{X}$, showing an ``X''-shaped dispersion at $\bar{X}$ at 2.4 eV. We show in Fig. 1(d) the bulk-band structure obtained by the first-principles band calculations along the ${\Gamma}X$ ($k_z$=0) and $ZR$ ($k_z$=$\pi$) lines \cite{XuPRB2015}. One finds a good agreement with the ARPES results in Fig. 1(c). In particular, the X-shaped band at $\bar{X}$ is well reproduced in the calculated band for the $ZR$ cut. The doubling of e2 band observed by ARPES [Fig. 1(c)] is well understood in terms of superposition of two bands in the ${\Gamma}X$ and $ZR$ lines [see Fig. 1(d)]; this suggests that, even though the ARPES spectrum in principle reflects the electronic states integrated over a wide $k_z$ region of bulk BZ, in reality the electronic states at $k_z$=0 and $\pi$ have dominant contributions \cite{KumiPRB1998}. The nearly-flat band at $\sim$2.0 eV (s1) seen in Fig. 1(c) has no counterpart in the calculation [Fig. 1(d)], suggesting that it is of SS origin.

\begin{figure}
 \includegraphics[width=3.3in]{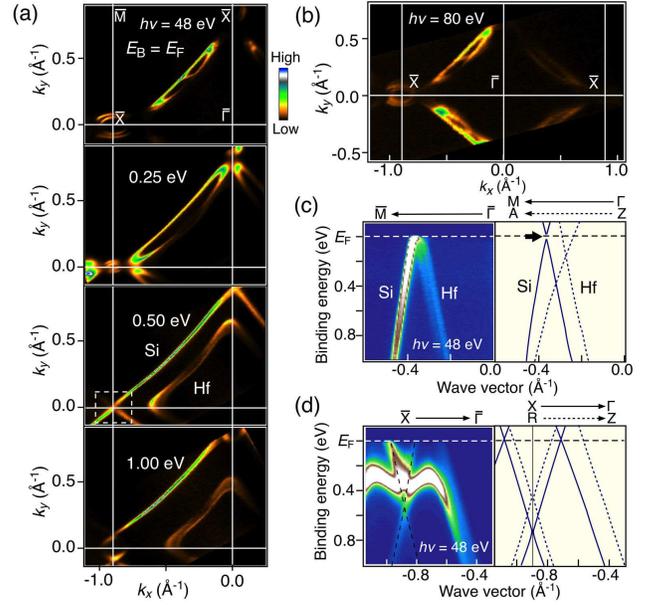}
  \hspace{-0.2in}
\caption{(color online). (a) ARPES-intensity mapping as a function of 2D wave vector for various $E_{\rm B}$'s from 0.0 eV ($E_{\rm F}$) to 1.0 eV, measured at $h\nu$ = 48 eV. (b) ARPES-intensity mapping at $E_{\rm F}$ at $h\nu$ = 80 eV. (c) Comparison of ARPES intensity along the $\bar{\Gamma}\bar{M}$ cut ($h\nu$ = 48 eV) and corresponding theoretical bulk-band dispersion \cite{XuPRB2015} for $k_z$=0 (solid curves) and $\pi$ (dashed curves). Arrow indicates the hybridization gap. (d) Same as (c) but along the $\bar{\Gamma}\bar{X}$ cut.}
\end{figure}

   To see more precise electronic states near the Fermi level ($E_{\rm F}$), we plot in Fig. 2(a) the ARPES intensity as a function of two-dimensional (2D) wave vector at several $E_{\rm B}$ slices measured at $h\nu$ = 48 eV. At $E_{\rm B}$ = $E_{\rm F}$, one finds a ``banana"-shaped FS elongated along the $\bar{X}\bar{X}$ direction, together with small pockets at $\bar{X}$. Similar FSs are also observed at $h\nu$ = 80 eV [Fig. 2(b)], suggesting that the electronic structure of HfSiS is quasi 2D as in ZrSiS \cite{AstNC2016, HasanPRB2016}. As seen in Fig. 2(a), the banana-like feature gradually expands upon increasing $E_{\rm B}$, and evolves into two diamonds at $E_{\rm B}$ $\geq$ 0.5 eV, whereas the small pockets at $\bar{X}$ gradually shrink and finally disappear. The good agreement of band dispersions along the $\bar{\Gamma}\bar{M}$ cut between experiments and calculations \cite{XuPRB2015} as shown in Fig. 2(c) suggests that the outer and inner diamonds in Fig. 2(a) arise from the holelike Si 3$p$ and the electronlike Hf 5$d$ bands at ${\Gamma}$, respectively. According to the band calculation, these bands intersect each other along the ${\Gamma}M$ or $ZA$ line to form line nodes in the case of negligible SOC \cite{XuPRB2015}. In the case of a finite SOC, a small hybridization gap opens along the ${\Gamma}M$ line. This small gap is not observed in the present study, because our crystal is lightly hole-doped so that the line nodes are lifted up into the unoccupied region \cite{AstNC2016}. 

As seen in Fig. 2(a), the outer diamond on the first and second BZs hits the $\bar{X}$ point at $E_{\rm B}$=0.5 eV, producing a ``X''-shaped pattern as highlighted by dashed rectangle. This feature is responsible for the band crossing at $\bar{X}$ in the ARPES intensity along $\bar{\Gamma}\bar{X}$ shown in Fig. 2(d), and the crossing is protected by glide-mirror symmetry of the crystal \cite{XuPRB2015, AstNC2016, HasanPRB2016}. The band calculation reproduces well the experimental band crossing at $\bar{X}$ [Fig. 2(d)], whereas the m-shaped feature at $E_{\rm B}${$\sim$}0.3 eV in the experiment has no counterpart in the calculation, suggesting its SS origin. These features of HfSiS are very similar to those of ZrSiS \cite{AstNC2016, HasanPRB2016}, supporting the LNSM nature of HfSiS. In fact, the SS origin of the m-shaped feature has been documented in ZrSiS by employing slab calculations.

 Figure 3(a) shows the FS mapping around $\bar{X}$ at $h\nu$ = 48 eV. One recognizes that there exist two FS pockets elongated along the $\bar{\Gamma}\bar{X}$ line (note that the intensity on the $\bar{\Gamma}\bar{X}$ line is suppressed due to the photoelectron matrix-element effect). The ARPES intensity along $\bar{X}\bar{M}$ in Fig. 3(b) [cut A in Fig. 3(a)] signifies a pair of V-shaped bands. These bands are assigned to the SSs since they are located in the gap in the bulk-band projection [see right panel of Fig. 3(b)]. As visualized in the second derivative of the ARPES intensity along cut A in Fig. 3(c), the pair of V-shaped bands become degenerate at $\bar{X}$; on the other hand, along cut B which does not pass through $\bar{X}$, the degeneracy of the two bands on the $\bar{\Gamma}\bar{X}$ line is lifted by $\sim$0.1 eV [Fig. 3(d)]. Taking into account that the $\bar{X}$ point is a time-reversal invariant momentum (TRIM), the V-shaped bands likely originate from Rashba splitting caused by space-inversion-symmetry breaking at the surface. It should be noted that Rashba splitting is not seen in ZrSiS \cite{AstNC2016, HasanPRB2016} or ZrSnTe \cite{ZrSnTeHong}, probably because of the lighter atomic mass of Zr ($Z$=40) compared to that of Hf ($Z$=72) and resultant weaker SOC in Zr compounds.
 
 \begin{figure}
 \includegraphics[width=3.3in]{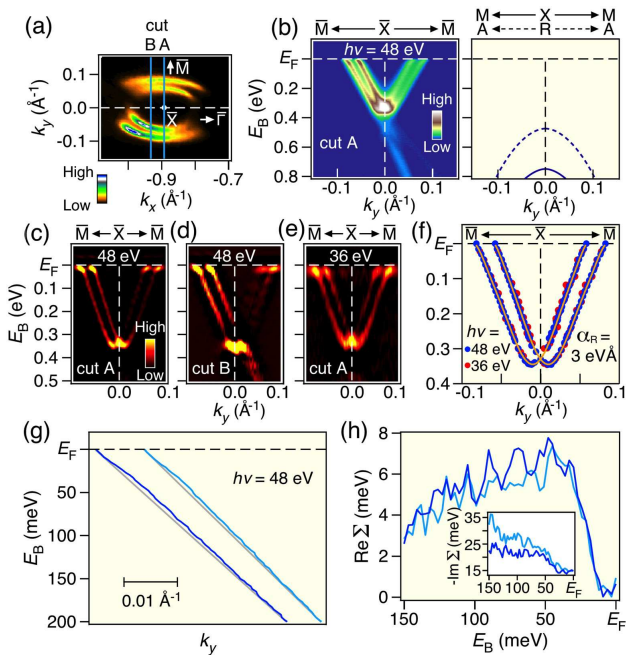}
 \hspace{-0.2in}
\caption{(color online). (a) ARPES-intensity mapping at $E_{\rm F}$ around $\bar{X}$ measured at $h\nu$ = 48 eV. Blue lines indicate the measured $k$ cuts. (b) Left: Near-$E_{\rm F}$ ARPES intensity along cut A ($\bar{X}\bar{M}$ cut). Right: theoretical bulk-band dispersions along $XM$ ($RA$) cut \cite{XuPRB2015} [solid (dashed) curves]. (c),(d) Second derivatives of the ARPES intensity along cuts A ($\bar{X}\bar{M}$ cut) and B (off-$\bar{X}\bar{M}$ cut), respectively, at $h\nu$ = 48 eV. (e) Same as (c) but measured at $h\nu$ = 36 eV. (f) Experimental band dispersion along $\bar{X}\bar{M}$ derived by tracing the peak position of EDCs at $h\nu$ = 48 and 36 eV. Orange curves represent the numerical fitting of the experimental band dispersion with two polynomial functions (up to 8th order was included) with $k$ offsets  $\pm{\Delta}k_{\rm R}$. (g) Peak position of MDCs for the Rashba SSs in the negative $k_y$ region (blue and light blue curves). Gray lines are the bare-band dispersion [same as orange curves in (f)]. (h) Real part of self-energy (Re$\Sigma$) estimated from the peak position of MDCs. Inset shows the imaginary part (-Im$\Sigma$) estimated from the peak width of MDCs.}
\end{figure}

 To evaluate the strength of the SOC, we extracted the experimental band dispersion for both $h\nu$ = 48 and 36 eV [Figs. 3(c) and 3(e)] by tracing the peak position of the energy distribution curves (EDCs) [circles in Fig. 3(f)]. The extracted dispersion, which is numerically fitted using two polynomial functions having opposite $k$ offsets $\pm{\Delta}k_{\rm R}$, yields the Rashba parameter ${\alpha}_{\rm R} = 2{\Delta}k_{\rm R}/{\Delta}E_{\rm R}$ = 3 eV{\AA}  (${\Delta}E_{\rm R}$ is the energy difference between the Kramers point and the band bottom \cite{TakayamaPRL2015}). This value is larger than those of typical Rashba SSs such as Au(111) (0.33 eV{\AA} \cite{LaShellPRL1996}) and Bi(111) (0.56 eV{\AA} \cite{KorotevPRL2004, TakayamaPRL2015}). The large Fermi velocity of 5.5 eV{\AA} in HfSiS may play an important role in the large ${\alpha}_{\rm R}$ value.
 
 \begin{figure*}
\begin{center}
\includegraphics[width=6.7in]{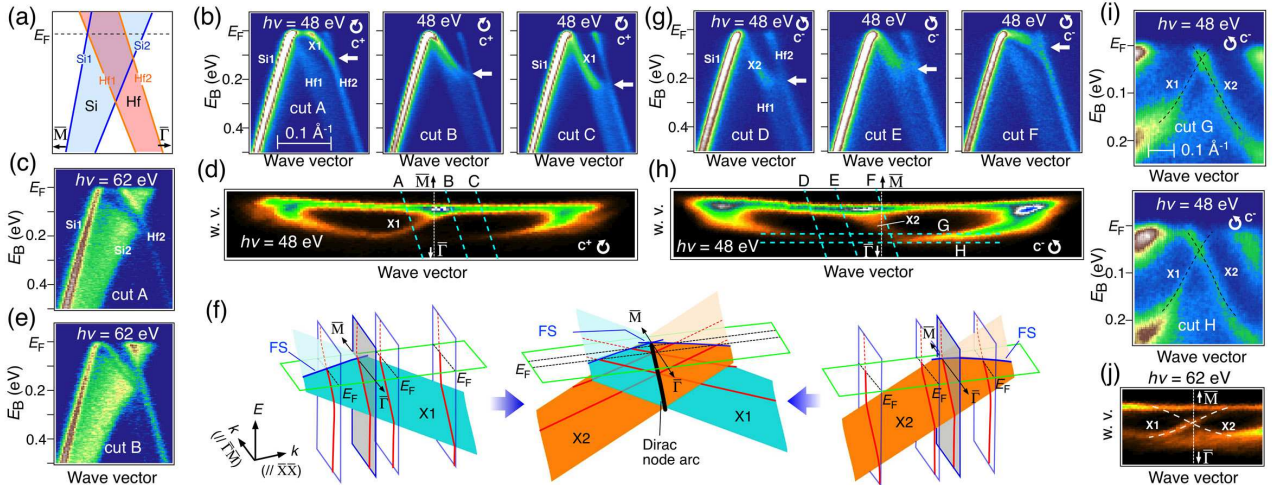}
\hspace{-0.2in}
\caption{(color online). (a) Schematic ARPES-intensity plot along $\bar{\Gamma}\bar{M}$ across the line node, which takes into account the calculated bulk-band dispersion and large $k_z$ broadening. The Si1 and Hf1 bands originate from $k_z$=0, while Si2 and Hf2 are from $k_z$=$\pi$. Shaded area indicates the bulk-band projection. (b) Plots of near-$E_{\rm F}$ ARPES intensity as a function of in-plane wave vector and $E_{\rm B}$ measured at $h\nu$ = 48 eV with right circular polarization (C$^+$) along cuts A-C shown in (d). (c) Near-$E_{\rm F}$ ARPES intensity along cut A measured at $h\nu$ = 62 eV. (d) Zoom of ARPES-intensity mapping at $E_{\rm F}$ around the banana-shaped FS measured at $h\nu$ = 48 eV with C$^+$ polarization. Blue dashed lines indicate the measured $k$ cuts. (e) Same as (c) but along cut B. (f) Schematic band dispersion in 3D $E$-$k$ space for (left) X1 band, (middle) X1+X2 bands, and (right) X2 band. Black line in the middle panel shows the Dirac-node arc. (g),(h) Same as (b),(d) but measured with left circular polarization C$^-$. (i) Near-$E_{\rm F}$ ARPES intensity along cuts G and H in (h). (j) Expanded view of the X-shaped FS at $h\nu$ = 62 eV (with C$^+$ polarization).}
 \end{center}
 \end{figure*}
 
  We found a signature of electron-mode coupling in the band dispersion. One can notice in Fig. 3(g) that the dispersion of the Rashba SSs derived by fitting the momentum distribution curves (MDCs) shows a small but finite deviation from the bare-band dispersion, exhibiting a weak kink at $E_{\rm B}$ $\sim$50 meV. Consequently, the real part of electron self-energy Re$\Sigma$ [Fig. 3(h)] shows a broad hump at $E_{\rm B}$ $\sim$50 meV, accompanied by a steep drop in the imaginary part $|{\rm Im}\Sigma|$ at this energy (inset). This feature arises from the coupling of surface electrons with a collective mode, most likely phonons originating from S and/or Si vibrations \cite{XuPRB2015}. The coupling constant estimated from the slope of Re$\Sigma$ is 0.2$\pm$0.05, comparable to those of typical SSs such as in Cu (0.16 \cite{EigurenPRB2002, TamaiPRB2013}), Ag (0.12 \cite{EigurenPRB2002}), and Bi (0.2-0.4 \cite{HofmannPSS2006}). 
  
  Now we present our most important finding, the {\it Dirac-node arc}. As schematically shown in Fig. 4(a), two Si (Si1, Si2) and two Hf (Hf1, Hf2) bulk bands (originating from $k_z$=0 and $\pi$ components) should delineate the bulk-band dispersions in the $\bar{\Gamma}$-$\bar{M}$ direction, with a weak intensity filling in-between (shaded area). By utilizing the selection rules of the photoelectron intensity, we were able to resolve all these features by tuning $h\nu$. For example, the ARPES intensity in Fig. 4(b) along cut A [see Fig. 4(d)] at $h\nu$ = 48 eV displays the two Hf bands (Hf1 and Hf2) and one Si band (Si1), while the two Si bands (Si1 and Si2) and one Hf band (Hf2) are seen at $h\nu$ = 62 eV [Figs. 4(c) and 4(e)]. Besides these bands, the ARPES intensity along cut A in Fig. 4(b) signifies an additional band (X1) which splits from the Hf2 band and displays a sudden change in the velocity at $E_{\rm B}$ = 0.1 eV (white arrow). This X1 band crosses $E_{\rm F}$ midway between the Fermi wave vectors ($k_{\rm F}$'s) of the Hf1 (as well as Si1) and Hf2 bands, and is responsible for forming a portion of the FS which traverses the top and bottom segments of the banana in Fig. 4(d). Upon moving away from the $\bar{\Gamma}\bar{M}$ line [from cut A to C in Fig. 4(b)], the X1 band gradually moves downward, and its $k_{\rm F}$ point merges into that of the Si1 band along cut C. This behavior is highlighted in the schematic 3D band plot in the left panel of Fig. 4(f) that illustrates the asymmetric X1 band with respect to the $\bar{\Gamma}\bar{M}$ line. Note that the X1 band exists only in the region confined between Si1 and Hf2.

Since the observed asymmetric dispersion apparently violates the mirror symmetry of the crystal, one has to find the counterpart with the dispersion symmetric to the X1 band with respect to the $\bar{\Gamma}\bar{M}$ line to preserve the mirror symmetry, as depicted in the right panel of Fig. 4(f). We were able to observe this band (X2) by switching the polarization of circularly polarized light to reverse the relative intensity between the two bands. As shown in Fig. 4(h), the switching of polarization of light leads to a change in the intensity distribution around $\bar{\Gamma}\bar{M}$. Consequently, the band dispersion in Fig. 4(g) (cuts D-F) follows the curve that is expected from the mirror reflection of the cuts A-C in Fig. 4(b). This suggests that the actual band has an X-shape tilted in $E$-$k$ space, resulting from a merger of X1 and X2 shown in the middle panel of Fig. 4(f). In fact, as shown in Fig. 4(i), such an X-shaped dispersion is observed when the $k$ cut is parallel to the banana [cut G in Fig. 4(h)]. We also found that the X-shaped band moves downward in cut H and the FS contains an X-shaped part [see the FS image at $h\nu$ = 62 eV in Fig. 4(j)]. These results led us to conclude that the X-shaped dispersion extends along a line on the $\bar{\Gamma}\bar{M}$ plane [black line in the middle panel of Fig. 4(f)], which can be viewed as an arc of Dirac node extending one-dimensionally in $k$-space \cite{AdamNC2016} and is confined between the bulk bands. This Dirac-node arc is apparently different from the Dirac cone of TIs and graphene where the upper and lower cones intersect at a point in $k$-space.

  The origin of the X1 and X2 bands to give rise to the Dirac-node arc is not clear. Since these bands split from the Hf2 band at the intersection of the Si2 and Hf2 bands [Figs. 4(c) and 4(e)], hybridization between the two bulk bands through SOC may play some role. However, X1 and X2 are not bulk bands, because they are not found in the bulk-band calculation \cite{XuPRB2015}. Also, the insensitivity of their dispersions to the photon energy supports their SS origin; however, they may not be well confined within the topmost layer due to the overlapping with the bulk-band projection. Obviously, these unexpected SSs pose a significant challenge in our understanding of topological LNSMs. In this regard, the appearance of the nearly-flat SS is understood to be a result of bulk-boundary correspondence associated with the topology of the bulk line nodes \cite{BurkovPRB2011}. The unexpected SSs containing a novel Dirac-node arc could also be a consequence of some topology which is yet to be discovered in LNSMs. Whatever its origin, the Dirac-node arc found here has a peculiar characteristic, that a charge neutrality point is always present in the SS as long as the arc crosses $E_{\rm F}$. 

In summary, we report ARPES results on HfSiS to show that this material is a line-node semimetal with the band crossing at $\bar{X}$ protected by glide-mirror symmetry. In addition, we observed (i) the SSs with a large Rashba splitting at $\bar{X}$, (ii) a kink in the SS dispersion at $\sim$50 meV, and (iii) unexpected X-shaped band dispersion with a nodal-arc structure. The novel Dirac-node arc is beyond the current understanding of topological line-node semimetals and poses an intriguing challenge to theory.

\begin{acknowledgements}
We thank N. Inami, K. Horiba, H. Kumigashira, and K. Ono for their assistance in ARPES measurements. This work was supported by MEXT of Japan (Innovative Area ``Topological Materials Science'', 15H05853), JSPS (KAKENHI 15H02105, 26287071, 25287079), KEK-PF (Proposal number: 2015S2-003), and DFG (Project A04 of CRC1238 ``Control and Dynamics of Quantum Matter'').
\end{acknowledgements}

%\clearpage

\bibliographystyle{prsty}

\newpage
{

\onecolumngrid
\begin{center}
{\large Supplemental Materials for \\
``Unexpected Dirac-Node Arc in the Topological Line-Node Semimetal HfSiS''}

\vspace{0.3 cm}

D. Takane,$^1$ Z. Wang,$^2$ S. Souma,$^{3,4}$ K. Nakayama,$^1$ C. X. Trang,$^1$ T. Sato,$^{1,4}$ T. Takahashi,$^{1,3,4}$ and Yoichi Ando$^2$

{\footnotesize
$^1${\it Department of Physics, Tohoku University, Sendai 980-8578, Japan}
\newline
$^2${\it Institute of Physics II, University of Cologne, K$\ddot{o}$ln 50937, Germany}
\newline
$^3${\it WPI Research Center, Advanced Institute for Materials Research,
Tohoku University, Sendai 980-8577, Japan}
\newline
$^4${\it Center for Spintronics Research Network, Tohoku University, Sendai 980-8577, Japan}
}

\end{center}

\twocolumngrid

\raggedbottom

\renewcommand{\thefigure}{S\arabic{figure}}
\setcounter{figure}{0}

\subsection{S1. Sample growth and characterization}
High-quality single crystals of HfSiS were synthesized with a chemical vapor transport method by using I$^2$ as transport agent. High-purity powders of Hf (99.9\%), Si (99.99\%) and S (99.999\%) were sealed in an evacuated quartz tube, which was put in a muffle furnace and was kept at 1100 $^{\circ}$C (source side) for 7 days with a temperature gradient of 10 $^{\circ}$C within the tube. To obtain high-quality crystals, we performed surface cleaning of Hf at 500 $^{\circ}$C using H$^2$ in a quartz tube [33]. Sample orientation was determined by Laue x-ray diffraction prior to the ARPES experiment. Typical Laue pattern from the [001] direction and the picture of single crystals are shown in Fig. S1 and its inset, respectively.

 \begin{figure}[h]
 \includegraphics[width=2 in]{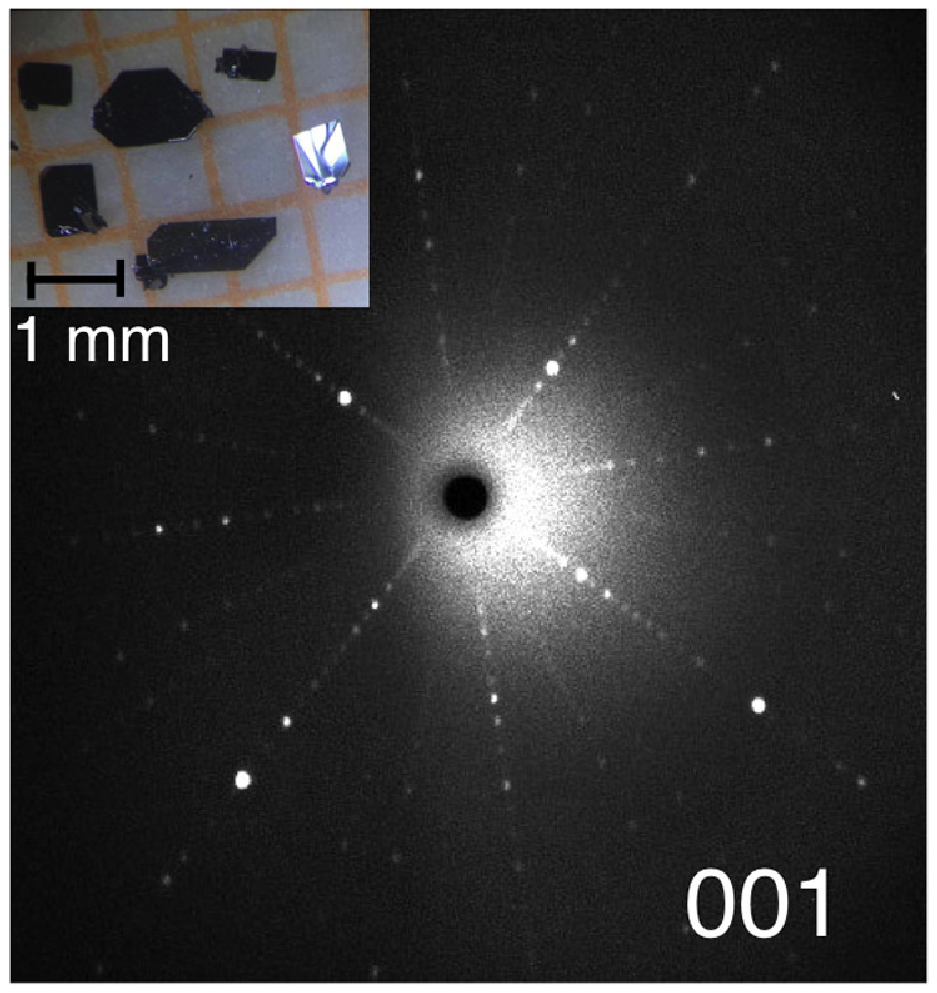}
\caption{Laue diffraction pattern from the [001] direction for HfSiS. Inset shows the picture of single crystals.
}
\end{figure}

\subsection{S2. ARPES experiments and core-level spectra}
 ARPES measurements were performed with an Omicron-Scienta SES2002 electron analyzer with energy-tunable synchrotron light at BL28A in Photon Factory. We used circularly polarized light of 36--200 eV. The energy and angular resolutions were set at 10--30 meV and 0.2$^{\circ}$, respectively. Samples were cleaved in situ along the (001) crystal plane in an ultrahigh vacuum of 1$\times$10$^{-10}$ Torr, and kept at 30 K during the measurements. Figure S2 displays the energy distribution curve (EDC) of HfSiS in a wide energy region measured at $h\nu$ = 200 eV. One can recognize sharp core-level features at the binding energy ($E_{\rm B}$) of about 165, 100, 50, and 20 eV, which are attributed to the S 2$p$, Si 2$p$, Hf 5$p$, and Hf 4$f$ orbitals, respectively. Sharp core-level peaks and absence of additional peaks in the EDC demonstrate high-quality nature of the cleaved surface.

\begin{figure}
 \includegraphics[width=3 in]{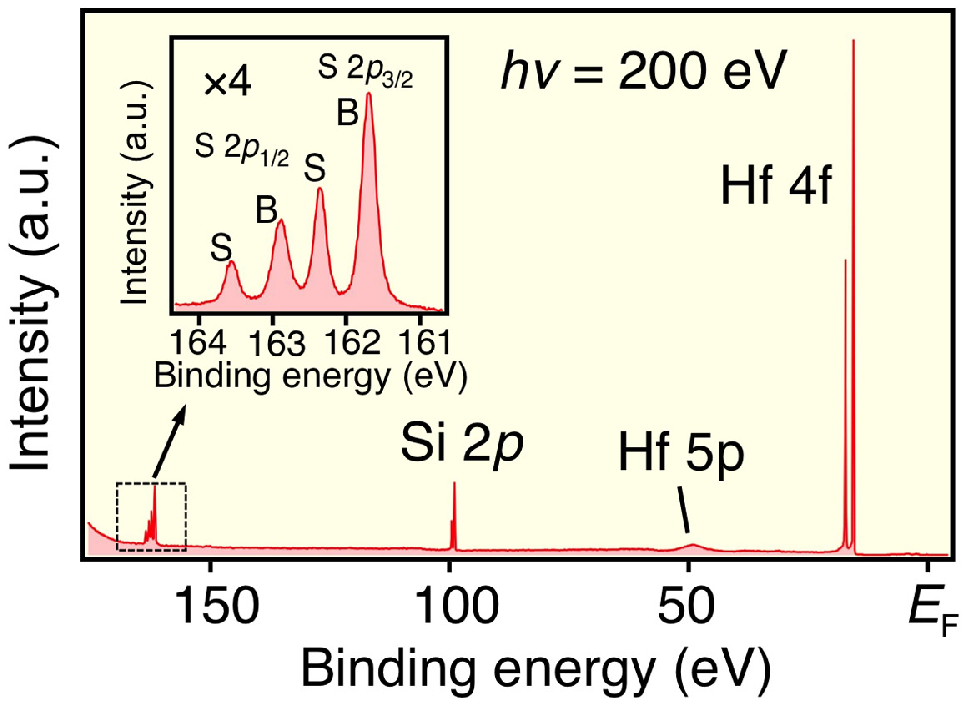}
\caption{ARPES spectrum of HfSiS in a wide energy region measured at $h\nu$ = 200 eV at 30 K. Inset shows the expansion of S 2$p$ core levels. ``S'' and ``B'' denote the surface and bulk components, respectively.
}
\end{figure}

\subsection {REFERENCES}

\noindent
[33] Z. Wang, K. Segawa, S. Sasaki, A. A. Taskin, and Y. Ando, APL Mater. \textbf{3}, 083302 (2015).

}

\end{document}